\documentclass[traditabstract,letter]{aa}  
\usepackage{epsfig}
\usepackage{natbib}
\usepackage{txfonts}
\usepackage{graphicx}

\def\m2s2{\hbox{\,m$^{2}$\,s$^{-2}$}} 

\def\Mjup{\hbox{$M_{\rm Jup}$}}

\def\Rear{\hbox{$R_{\oplus}$}\ }

\def\chisq{\mbox{$\chi^2$}\ }

\begin{document}

\title{The secondary eclipse of the transiting exoplanet CoRoT-2b \thanks{Based on observations obtained with CoRoT, a space project operated by the French Space Agency, CNES, with participation of the Science Program of ESA, ESTEC/RSSD, Austria, Belgium, Brazil, Germany, and Spain.} }

\author{Alonso, R. \inst{1}
\and Guillot, T. \inst{2}
\and Mazeh, T. \inst{3,4}
\and Aigrain, S. \inst{5}
\and Alapini, A. \inst{5}
\and Barge, P. \inst{6}
\and Hatzes, A. \inst{7}
\and Pont, F. \inst{5}
}

\offprints{\email{roi.alonso@unige.ch}}

\institute{Observatoire de Gen\`eve, Universit\'e de Gen\`eve, 51 Ch. des Maillettes, 1290 Sauverny, Switzerland
\and
Observatoire de la C\^ote d'Azur, Laboratoire Cassiop\'ee, CNRS UMR 6202, BP 4229, 06304 Nice Cedex 4, France
\and
Radcliffe Institute for Advanced Studies at Harvard, 8 Garden
St. Cambridge MA 02138 USA
\and
 School of Physics and Astronomy, R. and B. Sackler Faculty of Exact Sciences, Tel Aviv University, Tel Aviv 69978, Israel
\and
School of Physics, University of Exeter, Stocker Road, Exeter EX4 4QL, United Kingdom
\and
Laboratoire d'Astrophysique de Marseille, UMR 6110, Technopole de Marseille-Etoile,F-13388 Marseille cedex 13, France
\and
Th\"uringer Landessternwarte Tautenburg, Sternwarte 5, 07778 Tautenburg, Germany
}

\date{Received / Accepted }

\abstract{We present a study of the light curve of the transiting
exoplanet CoRoT-2b, aimed at detecting the secondary eclipse and measuring
its depth. The data were obtained
with the CoRoT satellite during its first run of more than
140~days. After filtering the low frequencies with a pre-whitening
technique, we detect a 0.0060$\pm$0.0020\% secondary eclipse centered on the
orbital phase 0.494$\pm$0.006. Assuming a black-body emission of the
planet, we estimate a surface brightness temperature of
T$_{\rm p,CoRoT}$=1910$^{+90}_{-100}$~K. We provide the planet's equilibrium temperature and re-distribution factors as a function of the unknown amount of reflected light. The upper limit for the geometric albedo is 0.12. The detected secondary is the shallowest ever found.}


\titlerunning{The secondary eclipse of CoRoT-2b}

\authorrunning{Alonso et al.}

\maketitle

\section{Introduction}
\label{sec:intro}

Out of the circa 350 exoplanets found to date, the 50 that transit
their host stars allow several follow-up studies, based either on the occultation of part of the
stellar disk by the planet (transits) or on the disappearance of the
planet behind the star (secondary eclipses). Observations of the
secondary eclipses of several exoplanets have enabled measurements of their
emitted flux at different wavelengths, providing clues to their
atmospheres (\citealt{char05,dem05}). Temperature inversions at stratospheric levels have been
found on several occasions among the hottest known
transiting Jupiters (\citealt{burr07,knu08,knu09}), and several detections
of the modulation of the planetary flux with orbital phase have been claimed by \cite{knu07} and
\cite{sne09}.

The transiting planet CoRoT-2b \citep{alo08a} was discovered during
the first $\sim$150-d pointing of the CoRoT mission
\citep{baglin06}. The spectral type of the star (G0V) and the period
of the planet (1.7\,d) suggest that it belongs to the $pM$ class of
exoplanets as defined by \cite{fort}, for which the strong incident
flux is thought to cause stratospheric thermal inversions. 

One of the intriguing characteristics of the hot Jupiters is their
variety in planetary radius. While for several planets the measured
radius agrees with theoretical models of planetary formation and
evolution, some of them appear larger than expected. A range of
scenarios have been suggested to account for the phenomenon, but no
clear picture has been delineated to date. CoRoT-2b not only
belongs to this class of ``bloated" planets, but it is one with the
most difficult radius as to explain, because of its high mass of 3.1\Mjup.

The host star of CoRoT-2b exhibits several signs of youth
\citep{bou08}, namely the detection of the Li {\sc I} 6708 \AA\ absorption line, an
inversion of the cores of the Ca{\sc II} $H$ and $K$ lines, and a
faster-than-expected rotation period of 4.5~d \citep{lanza09}. While these
``patterns" are traditionally attributed to stellar youth, it is also
possible that the stellar evolution has been affected by tidal
interactions with the close-in massive planet. Recently, \cite{jack09}
and \cite{pont09} have suggested some observational evidence for
strong tidal effects on exoplanets' host stars.

The activity of the star leaves footprints in the photometric light curve of CoRoT-2 that can be used to infer characteristics of the distribution and lifetimes of the spots on the stellar surface. By modeling the flux of the star in the parts of the light curve without transits, \cite{lanza09} observed a cyclic oscillation of the total spotted area of the star with a period of 28.9$\pm$4.3~days. In a different study, \cite{valio} looked at the effects of the occultation of spots in the planet's path along the stellar surface. 

In this paper we describe detection of the secondary eclipse in
the white light curve of the CoRoT public data. The technique is
similar to the one used to detect the secondary eclipse of CoRoT-1b
\citep{alo09b}. A tentative detection of a 5.5$\times$10$^{-5}$
eclipse, based on the same data, has been claimed by
\cite{alo08b}. Here we refine the analysis by carrying out a more careful filtering of the stellar variability, estimate the significance
of the detection, and discuss the results.

\section{Observations}
\label{sec:obs}

CoRoT-2b was observed during the first \emph{Long Run} pointing of the
satellite, which lasted 142-days. We used the data corrected to
the \emph{N2} level (the processing steps are described in \citealt{auv09}), which contains 369695 flux measurements, with a
time sampling of 512-s from the first 5.2 days of data, after which it
was changed to 32-s. The standard deviation of the normalized data
after filtering the low frequencies as described below is 0.0014 (in units of normalized flux) for the
32-s sampled data, while it is 0.00056 for the 512~s data. For
comparison, the photon noise level in the 32-s data is of 0.0011,
revealing that the real data is only about 20\% above the photon noise
level, increasing to 90\% above the photon noise in the 512-s part. An
inspection of the 2MASS plates reveals that the star has
a close companion 3.5~mags fainter in $V$, whose flux falls completely inside the aperture
mask. We therefore subtracted a constant value of 5.6\% of the median
level of the curve before normalizing it, as in
\cite{alo08a}.

\section{Analysis of the CoRoT light curve}
\label{sec:ana}

The two most remarkable features of the light curve of CoRoT-2 are the
transits by its companion and the modulation caused by the stellar
activity and rotation. Both features tend to hide any secondary
eclipse in the light curve; therefore, in this section we explain
our efforts to filter these two features, while avoiding to dilute or
erase the signal of the secondary eclipse.

\begin{figure}[!th]
\begin{center}
\epsfig{file=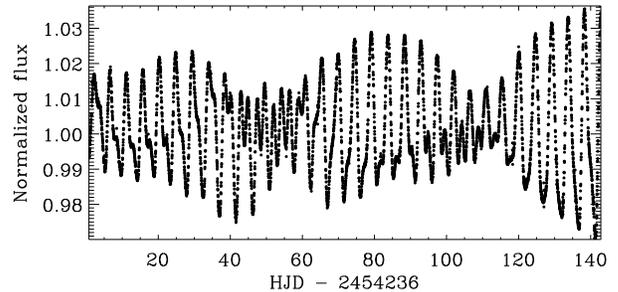,width=9cm}
\caption{Light curve of CoRoT-2 where the modeled transits have been subtracted. For display purposes, the data have been binned in 30~min bins.}
\label{fig:curve}
\end{center}
\end{figure}

To filter out the primary transit we could ignore the data points
inside the transit. However, inclusion of data gaps at the orbital period of
the transiting planet might result in spurious signals at this period
and its harmonics, risking an effect on the signal 
of secondary eclipse. To reduce this risk, we subtracted
the best fit solution of \cite{alo08a} from the curve, instead of cutting the
parts of the light curves where the transits are detected. We plot the light
curve with the transits removed this way in Fig.~\ref{fig:curve}. As
noted by \cite{valio}, the residuals in the individual transits show
the effect of occulted spots during the transit, so they might
add some high frequencies in a very short phase of the planet's
orbit.

To filter out the stellar activity signal, we 
pre-whitened the light curve using {\sc period04} \citep{lenz05}, a well-known
and tested technique for identifying periodic signals in variable
stars. Basically, it consists of successively locating the highest
amplitude in the Fourier power spectrum and fitting a combination of
sinusoids to the data that contain the frequency with the highest
amplitude and all the previously identified frequencies. Before
performing the {\sc period04} pre-whitening of the data, we averaged them
into 30-min bins. We searched for frequencies between 0 and 24~c/d (cycles per day).

\begin{figure}[!th]
\begin{center}
\epsfig{file=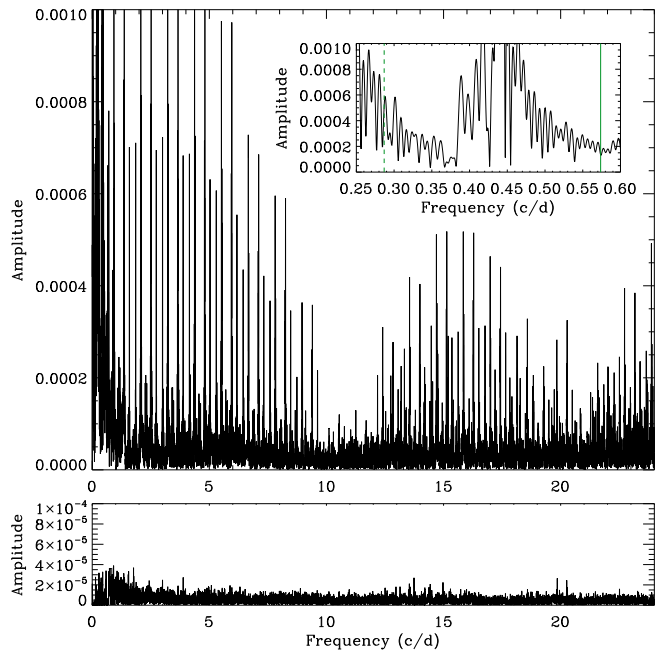,width=9cm}
\caption{The amplitude spectra of the CoRoT-2 light curve, before (top) and after (bottom) pre-whitening of the 123 frequencies needed to model the low frequency content of the signal. To avoid including gaps in the data, the transits were previously corrected by the best-fit model of \cite{alo08b}. The inset shows a zoom around the planet's orbital frequencies. The vertical solid and dashed green lines mark the planet's period and twice that value, respectively.}
\label{fig:fig2}
\end{center}
\end{figure}

The amplitude spectra before and after filtering the low
frequencies are plotted in Fig.~\ref{fig:fig2}, and the mean noise
level in the 0-24~c/d range in the filtered spectum is of
6.1$\times$10$^{-6}$, with a dispersion of 4.2$\times$10$^{-6}$.

The search for the secondary eclipse was performed using the same
techniques as described in \cite{alo09b}. Briefly, the search consisted of:
\begin{itemize}
\item 
Selecting an orbital phase $\phi_i$ where the secondary depth is
going to be evaluated.
\item 
Performing low order polynomial fits (linear fits in this particular case) to the parts of the light
curve surrounding the phase $\phi_i$. The regions where the fits are
evaluated are at a distance of slightly more than the transit
duration, to avoid including an hypothetical secondary centered
on $\phi_i$ in the fits. The total number of fits will thus be the
same as the total number of observed secondary eclipses.
\item 
Fitting a  trapezoid with the same duration and shape as obtained from
the transit parameters, and centered on $\phi_i$. The only free
parameter is the trapezoid depth.
\end{itemize}

\begin{figure}[!th]
\begin{center}
\epsfig{file=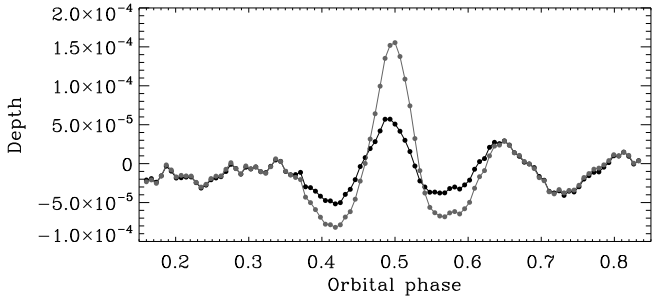,width=9cm}
\caption{Depth of a trapezoid with the shape and duration of a
  transit, as a function of the planet's orbital phase. The maximum is
  centered on phase 0.5, corresponding to the secondary eclipse. In
  gray, the result in a light curve where a secondary eclipse of 1$\times$10$^{-4}$ has been added to the raw light curve.}
\label{fig:fig_phase}
\end{center}
\end{figure}

The resulting plot of the significance and depth of a secondary
eclipse as a function of the orbital phase is given in
Fig.~\ref{fig:fig_phase}. The highest peak in this diagram appears
close to phase 0.5, corresponding to a depth of about
6$\times$10$^{-5}$.  

To make sure that any of the corrections explained above did not add or
remove some signal of the secondary eclipse, we inserted a 0.01\%
secondary eclipse in phase 0.5 to the raw light curve, and performed
the same analysis on this \emph{simulated} data.  The whole procedure,
using the same filtering and repeating the pre-whitening of the curve,
was performed on the \emph{simulated} data. The results are plotted in
Fig.~\ref{fig:fig_phase}, along with the results of the real
data. The highest peak in this case is also centered at 0.5, and the
fitted depth is of about 1.6$\times$10$^{-5}$, i.e., the sum of the
inserted eclipse and the signal present in the data. We thus suggest
that the whole preparation of the curve before performing the
secondary search does not significantly alter the signal we want to
measure.

\begin{figure}[!th]
\begin{center}
\epsfig{file=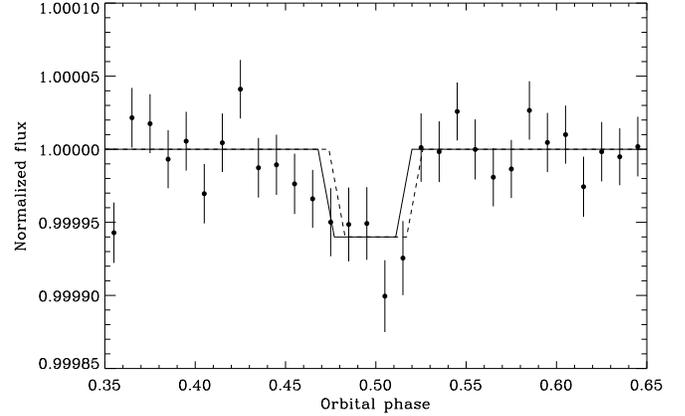,width=9cm}
  \caption{Phase-folded curve of CoRoT-2b during the phases of
    secondary eclipse. The data have been binned in 0.01 in phase
    ($\sim$25~min), and the 1-$\sigma$ error bars in each bin calculated as the standard deviation of the data points inside the bin divided by the square root of the number of points inside the bin. The solid line shows our best-fit trapezoid, with a depth of 0.0060$\pm$0.0020\%, and centered on phase 0.494$\pm$0.006. The dashed line shows a secondary eclipse centered on phase 0.5.}
\label{fig:fig4}
\end{center}
\end{figure}

\begin{figure}[!th]
\begin{center}
 \epsfig{file=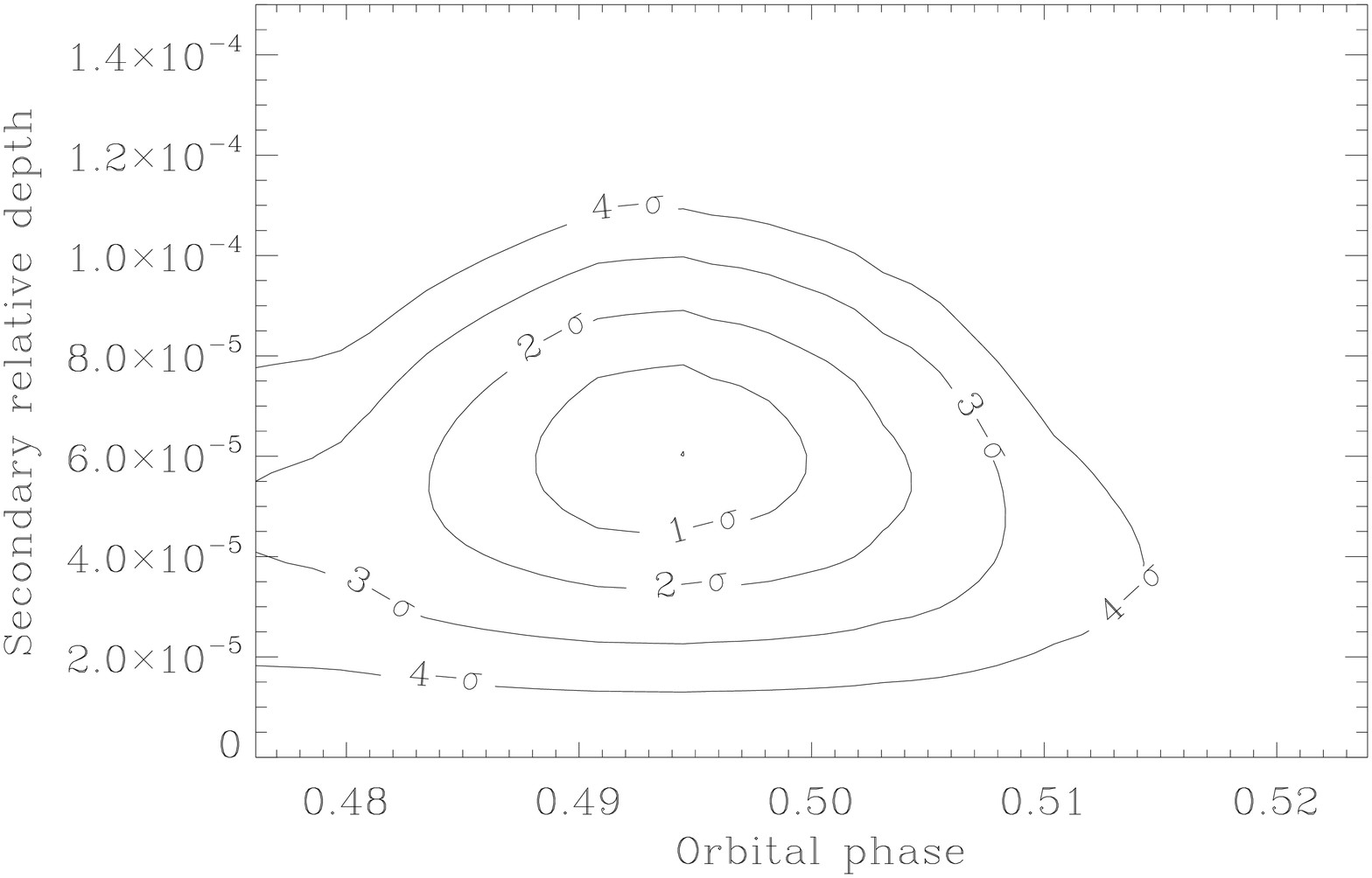,width=9cm}
  \caption{The \chisq value for different centers and depths of the
    secondary eclipse, and the 1,2,3, and 4-$\sigma$ confidence limits.}
\label{fig:chisq}
\end{center}
\end{figure}

The phase-folded light curve during the phases of expected secondary
eclipse is plotted in Fig.~\ref{fig:fig4}, together with the best
fitted trapezoids. We explored the goodness-of-fit through \chisq
function around phase 0.5, in a grid of secondary eclipse centers and
depths. The model used for the secondary was a trapezoid with
duration and shape fixed to those of the transit. The observed
secondary eclipse was binned into bins of 0.001 in phase
($\sim$2.5~min), and we estimated the error of each bin as the
standard deviation of the measurements inside the bin divided by the
square root of the number of points inside the bin. We plot the
resulting \chisq dependence and the formal sigma contours in
Fig.~\ref{fig:chisq}. A secondary eclipse with a depth of 0.0060$\pm$0.0020\% is detected in
the CoRoT light curve at an orbital phase of 0.494$\pm$0.006.

To estimate the significance of the secondary eclipse detection, we performed the following test, which preserved the presence of correlated noise in the data. From the data we subtracted the fitted secondary eclipse and shifted the residuals circularly, and then fitted a trapezoid with the phase and shape expected from the primary eclipse. Again, the only parameter was the secondary eclipse depth. All our 1000 trials resulted in depths less than 0.006\%, the depth derived from the real data. The mean fitted depth was 0.0001$\pm$0.0008\%, consistent with zero.
Additionally, we checked the uncertainties in the depth measurement by two different methods. In the first one, we reinserted the fitted secondary eclipse after the residual curves of the previous test were constructed and evaluated the fitted depths. We estimated the 1-$\sigma$ uncertainty from the standard deviation of the depths that were fitted during this test. The result of this test gave a depth of 0.0059$\pm$0.0008\%. 
As a second method, we fitted two Gaussian functions to (1) the distributions of points inside total secondary eclipse and (2) a subset of the part outside secondary eclipse, with the same number of points as (1) and with a randomly selected starting point. In the construction of the distributions, to minimize the effect of an arbitrarily selected histogram bin size, we used random values of the sizes of the histogram bin, between 1$\times$10$^{-6}$ and 7$\times$10$^{-6}$ in units of normalized flux. We repeated this test using 500 different starting points and values of the bin, and we estimated the 1-$\sigma$ errors as above. This test resulted in a depth of 0.0066$\pm$0.0020\%. We adopted a final value of the secondary eclipse depth of 0.006$\pm$0.002\% from the results of the different tests. 
 
The secondary eclipse appears slightly offset, at a 1-$\sigma$ level,
from the expected center for a circular orbit, which could be interpreted as a
possible non-zero eccentricity of $e\cos\omega$=0.01$\pm$0.01. This eccentricity is within the 1-$\sigma$ uncertainty of the orbital solution obtained from the radial velocity measurements (0.03$\pm$0.03 in \citealt{alo08a}).

\section{Discussion}
\label{sec:dis}

\begin{figure}[!th]
\begin{center}
 \epsfig{file=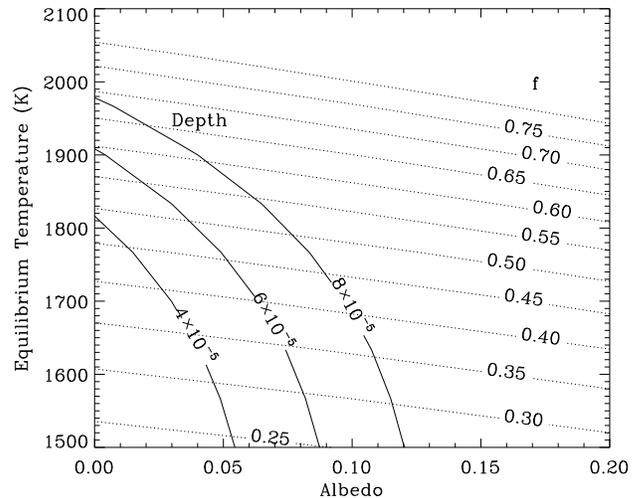,width=9cm}
\caption{The equilibrium temperature of CoRoT-2b as a function of the albedo matching the observed secondary eclipse depth. The zero-albedo equilibrium temperature is about 1910~K. The dotted lines plot the re-distribution factors $f$. Realistic re-distribution factors require the contribution of some reflected light (non-zero albedo) or the presence of an additional source of energy in the planet (assumed zero in this plot).}
 \label{fig:fig6}
\end{center}
\end{figure}

In the optical, the observed flux of the planet can be expressed:
\begin{equation}
F_{\rm p} = F_{\rm p,reflected} + F_{\rm p,reemitted} + F_{\rm p,internal}, 
\end{equation}
i.e., the sum of a component of reflected light, a thermal re-emission of the incident stellar flux, and a thermal emission not related to the incident flux from the star (e.g., internal heat or emission due to tidal forces).
The reflected light can be estimated as $F_{\rm p,reflected}=A_g\left(R_p\over a\right)^2$ where $A_g$ is the geometric albedo. 
We can calculate the thermal component of the planet's emission (the combination of $F_{\rm p,reemitted}$ and $F_{\rm p,internal}$) using the same procedure as in \cite{alo09b}. For that purpose, we used a value of T$_{\rm eff}$=5625$\pm$120~K for the star \citep{alo08a} and the model spectrum of a G8V star from \cite{pickles}. We calibrated the model spectrum to obtain the same integrated flux as a Planck function with the T$_{\rm eff}$ of the star. Taking the CoRoT spectral response function into account \citep{auv09}, the ratio of areas of the star and the planet, and the incident flux lost in the reflection ($A_{\rm B}  F_\star$ where $A_{\rm B}$  is
the Bond albedo), we searched for the brightness temperature that matched the observed eclipse depth. We assumed a black-body emission for the planet and considered different values of the albedo. We assumed $A_{\rm B}=A_{\rm g}$, a reasonable assumption since more than 50\% of the stellar flux is emitted at the CoRoT wavelengths. As an example, $A_{\rm B}\sim A_{\rm g}$ within 20\% for the four solar system giant planets when integrating $A_{\rm g}$ between 0.4 and 0.9$\,\mu$m \citep{kar94}. The different possible solutions are plotted in Figure~\ref{fig:fig6}. The zero-albedo brightness temperature in the CoRoT bandpass calculated this way and including the uncertainties on the T$_{\rm eff}$ resulted in T$_{\rm p,CoRoT}$=1910$^{+90}_{-100}$~K.  If we further assume that $F_{\rm p,internal}=0$ and that the planet is in thermal equilibrium\footnote{$T_{eq}=T_\star(R_\star/a)^{1/2}[f(1-A)]^{1/4} $}, this temperature favors high values of the re-distribution factor $f$=0.60$\pm$0.14 that are greater than the maximum expected for no re-distribution from the day to the night sides ($f=0.5$). This may be explained by (i) departures from a blackbody planetary emission, (ii) a non-zero albedo, or (iii) emission of significant internal energy from the planet. We defer the first possibility to future studies of radiative transfer in this planet. 

Assuming blackbody radiation and negligible $F_{\rm p,internal}$, Fig.~\ref{fig:fig6} shows that likely values of the albedo (corresponding to $f$ between 0.25 and 0.5) are $A_{\rm g}=0.06\pm 0.06$, which confirms the very low reflection of the planet obtained theoretically(e.g. \citealt{marley99,sud00,sea00,hood08}), and observationally: upper limits for the albedos of exoplanets in similar conditions as CoRoT-2b have been reported by Charbonneau et al. (1999, $A_g<$0.3 for $\tau$ Bootis b), Leight et al. (2003a,b, $A_g<$0.39 for $\tau$ Bootis b, $A_g<$0.12 for HD~75289), by \cite{rowe} on HD~209458b (a 3-$\sigma$ upper limit of 0.17). Finally, CoRoT-1b was found to be such that $A_{\rm g}<0.20$ using CoRoT's red channel of the light curve \citep{sne09}, and independently by \cite{alo09b} using the white channel. 

To distinguish among the different components of the observed planetary flux, we would need chromatic information. The CoRoT data are delivered in three band-passes, but we could not reach the level of precision needed to detect the secondary in these data. We attribute this to the difficulty of achieving a good correction of the jitter in the satellite's pointing.  

As a concluding remark, the depth of the secondary eclipse of CoRoT-2b is roughly the same as a transit of a 1\Rear planet in front of a solar radius star. The significance of the detection of the secondary eclipse thus emphasizes the excellent capabilities of the CoRoT mission at detecting planets with radii of only a few \Rear.

\begin{acknowledgements}
R.A acknowledges support by the grant CNES-COROT-070879. AH acknowledges the support of DLR grant 50OW0204

\end{acknowledgements}
%

\end{document}